# Scaled Atomic-Layer-Deposited Indium Oxide Nanometer Transistors with Maximum Drain Current Exceeding 2 A/mm at Drain Voltage of 0.7 V

Mengwei Si, Zehao Lin, Adam Charnas and Peide D. Ye

*School of Electrical and Computer Engineering and Birck Nanotechnology Center, Purdue University, West Lafayette, IN 47907, United States*

ABSTRACT

In this work, we demonstrate scaled back-end-of-line (BEOL) compatible indium oxide ($In_2O_3$) transistors by atomic layer deposition (ALD) with channel thickness ($T_{ch}$) of 1.0-1.5 nm, channel length ($L_{ch}$) down to 40 nm, and equivalent oxide thickness (EOT) of 2.1 nm, with record high drain current of 2.0 A/mm at $V_{DS}$ of 0.7 V among all oxide semiconductors. Enhancement-mode $In_2O_3$ transistors with $I_D$ over 1.0 A/mm at $V_{DS}$ of 1 V are also achieved by controlling the channel thickness down to 1.0 nm at atomic layer scale. Such high current density in a relatively low mobility amorphous oxide semiconductor is understood by the formation of high density 2D channel beyond $4\times10^{13}$ /$cm^2$ at $HfO_2/In_2O_3$ oxide/oxide interface.

KEYWORDS: indium oxide, wide bandgap, oxide semiconductor, thin-film transistor, charge neutrality level, ultrathin body, enhancement-mode

OXIDE semiconductors [1] are the leading channel materials for thin-film transistors (TFTs) and are considered as promising channel materials for back-end-of-line (BEOL) compatible transistors for monolithic 3-dimensional (3D) integration. Indium oxide ($In_2O_3$) [2] or doped $In_2O_3$ such as indium tin oxide (ITO) [3,4], indium tungsten oxide (IWO) [5], indium aluminum zinc oxide (IAZO) [6], indium gallium zinc oxide (IGZO), etc. [7-9], deposited by sputtering [3-9] or atomic layer deposition (ALD) [2,10-13], are being investigated due to their low thermal budget, high mobility, atomically smooth surface, wafer-scale homogenous films. Especially, the conformal capability of ALD on side walls, deep trenches, 3D structures enables tremendous new opportunities and the flexibility toward 3D device integration.

In this work, we report high-performance $In_2O_3$ transistors by ALD with channel length ($L_{ch}$) scaled down to 40 nm, with record high drain current ($I_D$) of 2.0 A/mm at a low drain-to-source voltage ($V_{DS}$) of 0.7 V, among all oxide semiconductors to the authors' best knowledge. Channel thickness ($T_{ch}$) scaling down to 1.0 nm is achieved by the accurate thickness control of ALD cycles. The devices exhibit excellent immunity to short channel effects (SCEs) due to $T_{ch}$ scaling and equivalent oxide thickness (EOT) scaling down to 2.1 nm. Enhancement-mode $In_2O_3$ transistors with threshold voltage ($V_T$) greater than zero and with $I_D$ over 1.0 A/mm at $V_{DS}$ of 1 V are also achieved by ALD control of thickness. Such high current density in a relatively low mobility amorphous oxide semiconductor is understood by the formation of high density 2D electron channel larger than $4 \times 10^{13}$ /cm$^2$ at $HfO_2/In_2O_3$ oxide/oxide interface [4].

Fig. 1(a) shows the schematic diagram of an $In_2O_3$ transistor. The gate stack includes 40 nm Ni as gate metal, 5 nm $HfO_2$ as gate dielectric, 1/1.2/1.5 nm $In_2O_3$ as semiconducting channels and 80 nm Ni as source/drain (S/D) contacts. Fig. 1(b) shows the scanning electron microscopy (SEM) image of a fabricated device with $L_{ch}$ of 1 μm, where the inset illustrates the measurement of the shortest $L_{ch}$ of 40 nm. $In_2O_3$ channel is too thin to be visible. $T_{ch}$ are determined together by transmission electron microscopy (TEM), atomic force microscopy (AFM) and ellipsometry, as shown in the $In_2O_3$ thicknesses versus ALD cycles in Fig. 1(c) [2]. The deposition rate is slower in the first 100 cycles due to the nucleation delay of a typical ALD process.

The device fabrication process started with standard cleaning of p+ Si substrate with 90 nm thermally grown $SiO_2$. A bi-layer photoresist lithography process was then applied for the sharp lift-off of Ni gate metal by e-beam evaporation. 5 nm $HfO_2$ was then deposited by ALD at 200 °C, using $[(CH_3)_2N]_4Hf$ (TDMAHf) and $H_2O$ as Hf and O precursors. $In_2O_3$ thin films with thicknesses of 1/1.2/1.5 nm were then deposited by ALD at 225 °C, using $(CH_3)_3In$ (TMIn) and $H_2O$ as In and O precursors. ALD was carried out using $N_2$ as carrier gas at a flow rate of 40 sccm and the base pressure is 432 mTorr. TMIn and $H_2O$ were pulsed for 625 ms and 750 ms at each cycle, respectively. $N_2$ flow rate was increased to 100 sccm during the 25 s purge. Channel isolation was done by wet etching of $In_2O_3$ using concentrated hydrochloric acid. 80 nm Ni was then deposited by e-beam evaporation as S/D contacts, patterned by electron beam lithography. The fabrication process has a low thermal budge of 225 °C and is BEOL compatible. The gate stack has an EOT of 2.1 nm as shown in the C-V measurement in Fig. 1(d). EOT is calculated using $C_{ox} = \frac{\epsilon_0 \epsilon_{SiO2}}{EOT}$, where $\epsilon_{SiO2}$ is 3.9 as dielectric constant of $SiO_2$, $\epsilon_0$ is $8.85 \times 10^{-14}$ F/cm as vacuum permittivity and $C_{ox}$ is measured from C-V measurement as 1.62 µF/cm².

Fig. 2(a) and 2(b) show the $I_D$-$V_{GS}$ and $I_D$-$V_{DS}$ characteristics of an $In_2O_3$ transistor with $L_{ch}$ of 40 nm and $T_{ch}$ of 1.2 nm. Maximum $I_D$ of 2.0 A/mm is achieved at a low $V_{DS}$ of 0.7 V. A low on resistance ($R_{ON}$) of 0.35 Ω·mm is obtained. Fig. 2(c) and 2(d) present the $I_D$-$V_{GS}$ and $I_D$-$V_{DS}$ characteristics of a similar $In_2O_3$ transistor with $L_{ch}$ of 50 nm and $T_{ch}$ of 1.2 nm. Maximum $I_D$ of 2.0 A/mm is achieved at $V_{DS}$ of 0.8 V. Fig. 2(e) and 2(f) illustrate the $I_D$-$V_{GS}$ and $I_D$-$V_{DS}$ characteristics of another In$_2$O$_3$ transistor with L$_{ch}$ of 1 μm and T$_{ch}$ of 1.2 nm, showing well-behaved I$_D$ saturation at high V$_{DS}$ greater than V$_{GS}$-V$_T$.

Fig. 3(a) and 3(b) show the I$_D$-V$_{GS}$ and I$_D$-V$_{DS}$ characteristics of an In$_2$O$_3$ transistor with L$_{ch}$ of 80 nm and T$_{ch}$ of 1.2 nm. Maximum I$_D$ of 2.1 A/mm is achieved at V$_{DS}$ of 1 V. V$_{GS}$-dependent extrinsic field-effect mobility (μ$_{FE}$) is extracted from maximum transconductance (g$_m$) at low V$_{DS}$, with a μ$_{FE}$ of 39 cm$^2$/V·s, as shown in Fig. 3(c). 2D carrier density (n$_{2D}$) can be *estimated* according to I$_D$ = n$_{2D}$qμE, where μ is mobility (V$_{GS}$-dependent μ$_{FE}$ is used) and E is the channel electric field (i.e. V$_{DS}$/L$_{ch}$ at low V$_{DS}$ assuming very low R$_C$), q is the elementary charge. A high 2D electron density at HfO$_2$/In$_2$O$_3$ oxide/oxide of 4.5×10$^{13}$ /cm$^2$ is achieved, suggesting Fermi level is deeply aligned into the conduction band (E$_C$) leading to high electron density and low contact resistance in In$_2$O$_3$ [2,4]. The high mobile carrier density is not screened by traps due to the Fermi level alignment inside of conduction band of In$_2$O$_3$.[2] The obtained high electron density is reasonable, considering on high gate capacitance of 1.6 μF/cm$^2$ (see Fig.1(d)), large voltage span > 4V, depletion-mode operation, a large bandgap of oxide channel.

Fig. 4(a) and 4(b) show the I$_D$-V$_{GS}$ and I$_D$-V$_{DS}$ characteristics of an In$_2$O$_3$ transistor with L$_{ch}$ of 40 nm and T$_{ch}$ of 1 nm. Maximum I$_D$ of 1 A/mm is achieved at V$_{DS}$ of 1 V. V$_T$ of 0.1 V is extracted by linear extrapolation at V$_{DS}$ of 0.05 V. Thus, enhancement-mode operation and high I$_D$ of 1 A/mm are achieved simultaneously. Fig. 4(c) and 4(d) show the I$_D$-V$_{GS}$ and I$_D$-V$_{DS}$ characteristics of an In$_2$O$_3$ transistor with T$_{ch}$ of 1.5 nm but L$_{ch}$ as large as 0.3 μm. Maximum I$_D$ of 1 A/mm is also achieved at V$_{DS}$ of 1 V, with a depletion-mode operation due to a relatively thick T$_{ch}$.

Fig. 5 summarizes the scaling metrics of In$_2$O$_3$ transistors with L$_{ch}$ from 1 μm down to 40 nm and with various T$_{ch}$ from 1.5 nm down to 1 nm. Each data point represents the average of at least 5 devices. The small error bar in these plots demonstrates that the ALD based $In_2O_3$ transistors are highly uniform. Fig. 5(a) and 5(b) show the maximum $I_D$ ($I_{D,max}$) and $g_m$ versus $L_{ch}$ characteristics at various $T_{ch}$. $I_{D,max}$ and $g_m$ are extracted at $V_{DS}=1$ V unless otherwise specified. The devices mostly follow a 1/L scaling trend. The deviation from 1/L scaling at short channel devices is because of lower $V_{DS}$ and self-heating effects. The deviation from 1/L scaling at long channels is likely to be the result of floating body effect. Fig. 5(c) studies the impact of $T_{ch}$ and $L_{ch}$ on $V_T$. Both depletion-mode and enhancement-mode $In_2O_3$ transistors are demonstrated. $V_T$ can be considerably tuned by $T_{ch}$ and accurately controlled by ALD cycles. Fig. 5(d) shows the scaling metrics of $In_2O_3$ transistors with various $T_{ch}$ on $\mu_{FE}$. $\mu_{FE}$ is extracted from maximum $g_m$ at low $V_{DS}$ of 0.05 V. High $\mu_{FE}$ of 77 $cm^2/V \cdot s$ is achieved at ultrathin $T_{ch}$ of 1.5 nm, which is rather high among amorphous oxide semiconductors, being benefitted from the atomically smooth surface by ALD. Fig. 5(e) presents the subthreshold slope (SS) versus $L_{ch}$ characteristics at high $V_{DS}$. Minimum SS of 88 mV/dec is achieved. SS has larger variation because off-state is more affected by gate leakage current, especially at short channel due to the more negative $V_T$. Such variation can be reduced by optimizing the gate stack. The devices exhibit excellent immunity to short channel effects down to 40 nm due to the ultrathin $In_2O_3$ channel and scaled EOT. The device performance has still rooms to boost by further aggressive scaling and process optimization.

Fig. 6 shows the TLM extraction of $R_C$ on $In_2O_3$ transistors with various $T_{ch}$ at constant $V_{GS}-V_T$. The y-axis intersection at $L_{ch}=0$ μm is extracted as $2R_C$. $R_C$ and contact resistivity ($\rho_C$) are calculated as shown in Fig. 6(a) and 6(b). $R_C$ as low as 0.06 $\Omega \cdot mm$ and $\rho_C$ as low as $0.5 \times 10^{-8}$ $\Omega \cdot cm^2$ are estimated on $In_2O_3$ transistors with $T_{ch}$ of 1.2 nm, indicating a very low effective Schottky barrier height and width.

In summary, scaled BEOL compatible ALD In$_2$O$_3$ transistors are demonstrated with T$_{ch}$ down to 1 nm, L$_{ch}$ down to 40 nm and EOT of 2.1 nm. A high I$_D$ of 2.0 A/mm at V$_{DS}$ of 0.7 V is achieved on depletion-mode In$_2$O$_3$ transistors. Enhancement-mode In$_2$O$_3$ transistors with I$_D$ over 1.0 A/mm at V$_{DS}$ of 1 V are also achieved, by ALD control of channel thickness on V$_T$ tuning. Such high current density in a relatively low mobility amorphous oxide semiconductor is understood by the formation of high density 2D electron density beyond $4\times10^{13}$ /cm$^2$ at HfO$_2$/In$_2$O$_3$ oxide/oxide interface. ALD In$_2$O$_3$ based devices are promising BEOL compatible device technology toward monolithic 3D integration. This new channel material at 1 nm atomic scale, as thin as monolayer of 2D van der Waals materials, opens tremendous new opportunities in device research.

The authors gratefully acknowledge X. Sun and H. Wang for the technical support on TEM imaging.

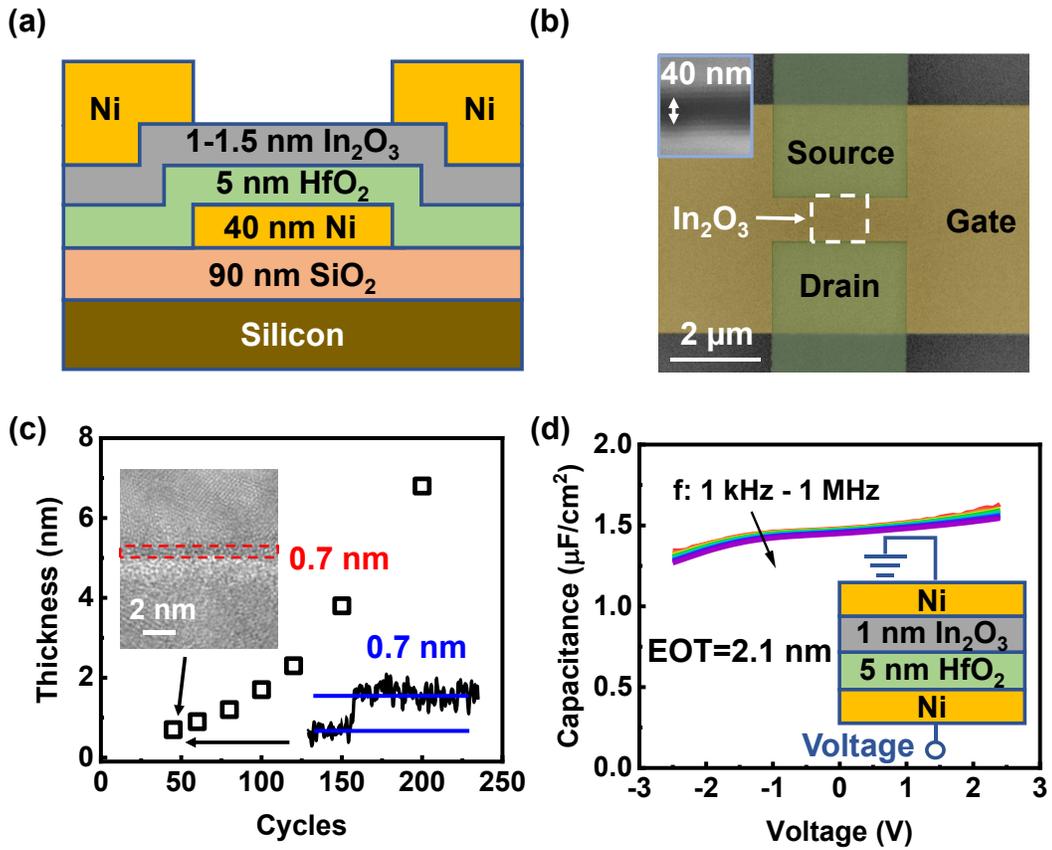

Fig. 1. (a) Schematic diagram of an In$_2$O$_3$ transistor with 5 nm HfO$_2$ as gate dielectric. (b) SEM image of a fabricated In$_2$O$_3$ transistor. (c) In$_2$O$_3$ thickness versus ALD cycles, showing a nucleation delay process. (d) C-V measurement of the gate stack from 1 kHz to 1 MHz. Smaller capacitance at negative gate bias is due to the depletion of semiconducting In$_2$O$_3$ channel.

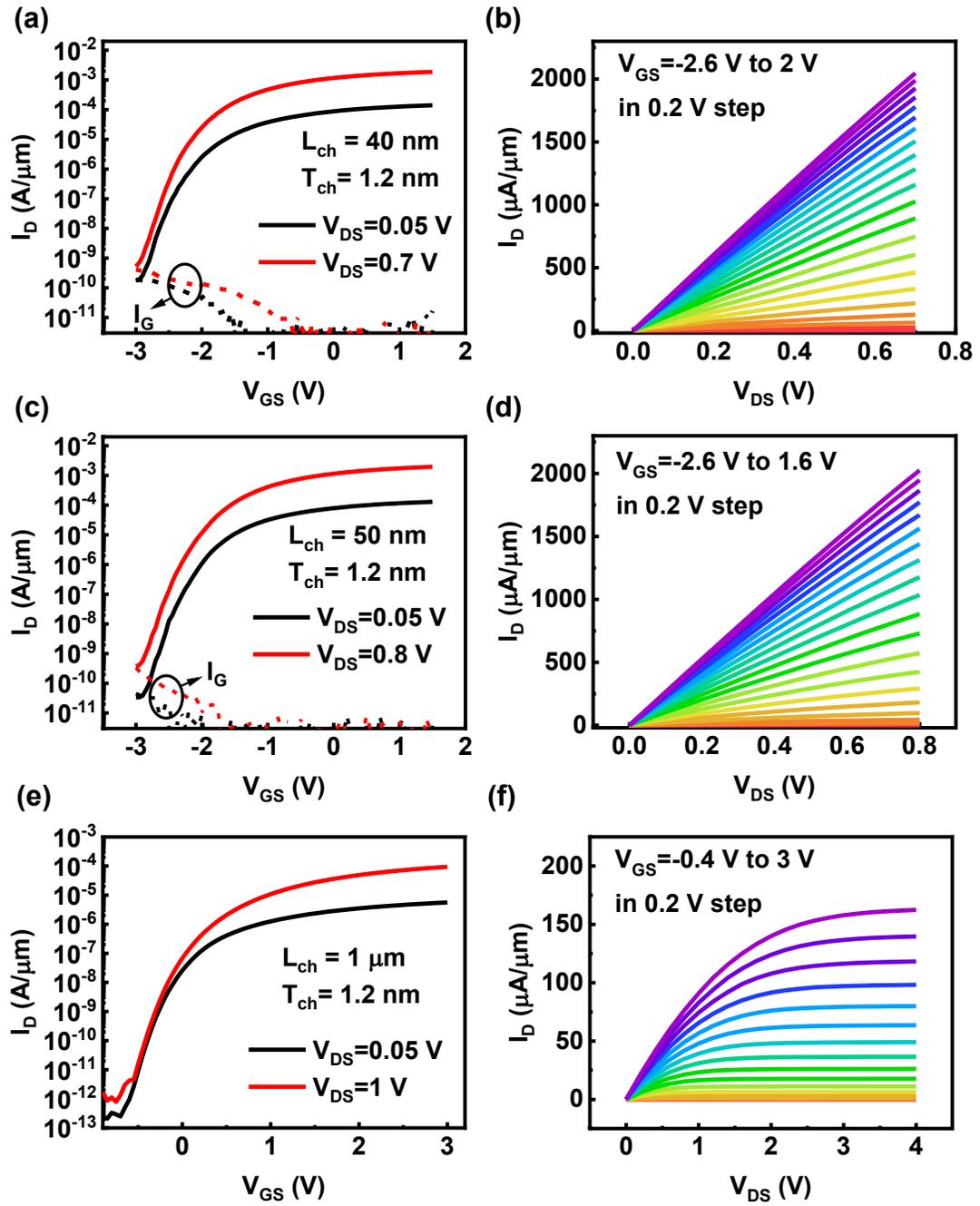

Fig. 2. $I_D$-$V_{GS}$ and $I_D$-$V_{DS}$ characteristics of $In_2O_3$ transistors with $L_{ch}$ of (a, b) 40 nm, (c, d) 50 nm, and (e, f) 1 μm and $T_{ch}$ of 1.2 nm.

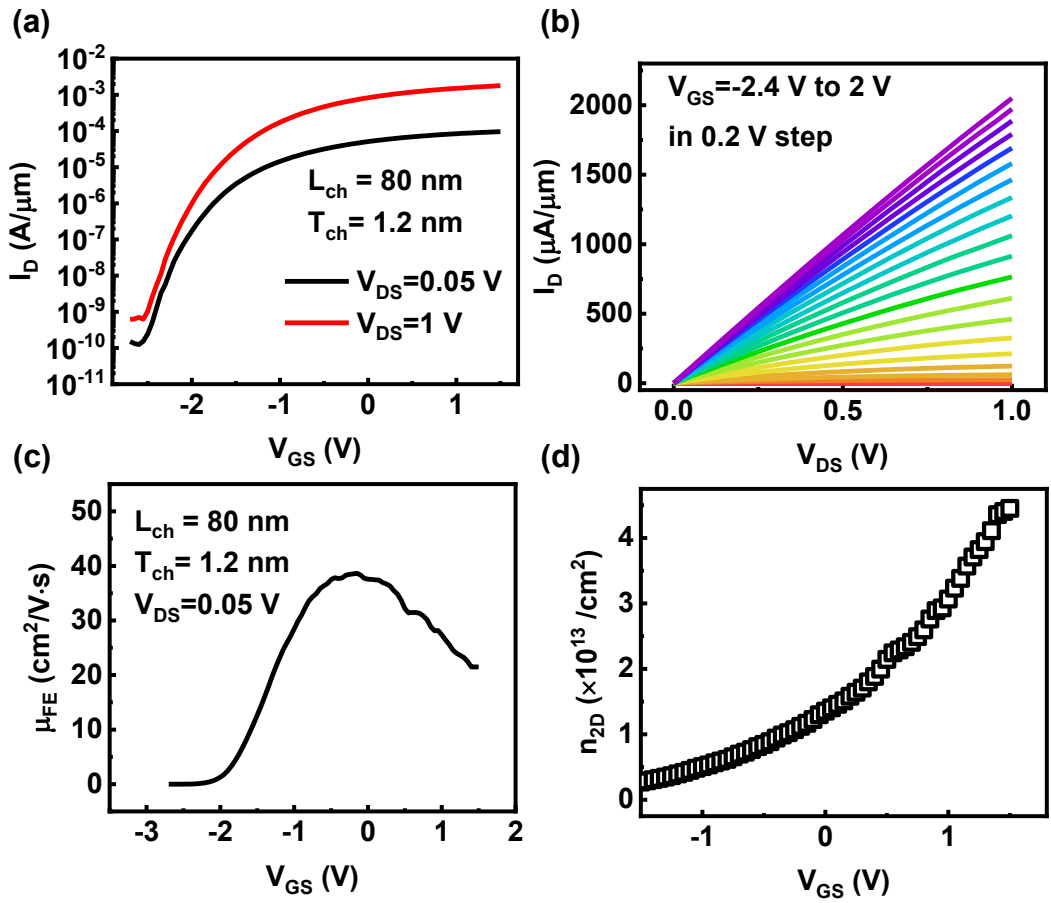

Fig. 3. (a) $I_D$-$V_{GS}$ and (b) $I_D$-$V_{DS}$ characteristics of an $In_2O_3$ transistor with $L_{ch}$ of 80 nm and $T_{ch}$ of 1.2 nm. (c) $\mu_{FE}$ versus $V_{GS}$ characteristics extracted at $V_{DS}$=0.05 V. (d) Channel mobile carrier density versus $V_{GS}$ calculated from $I_D$ and $\mu_{FE}$.

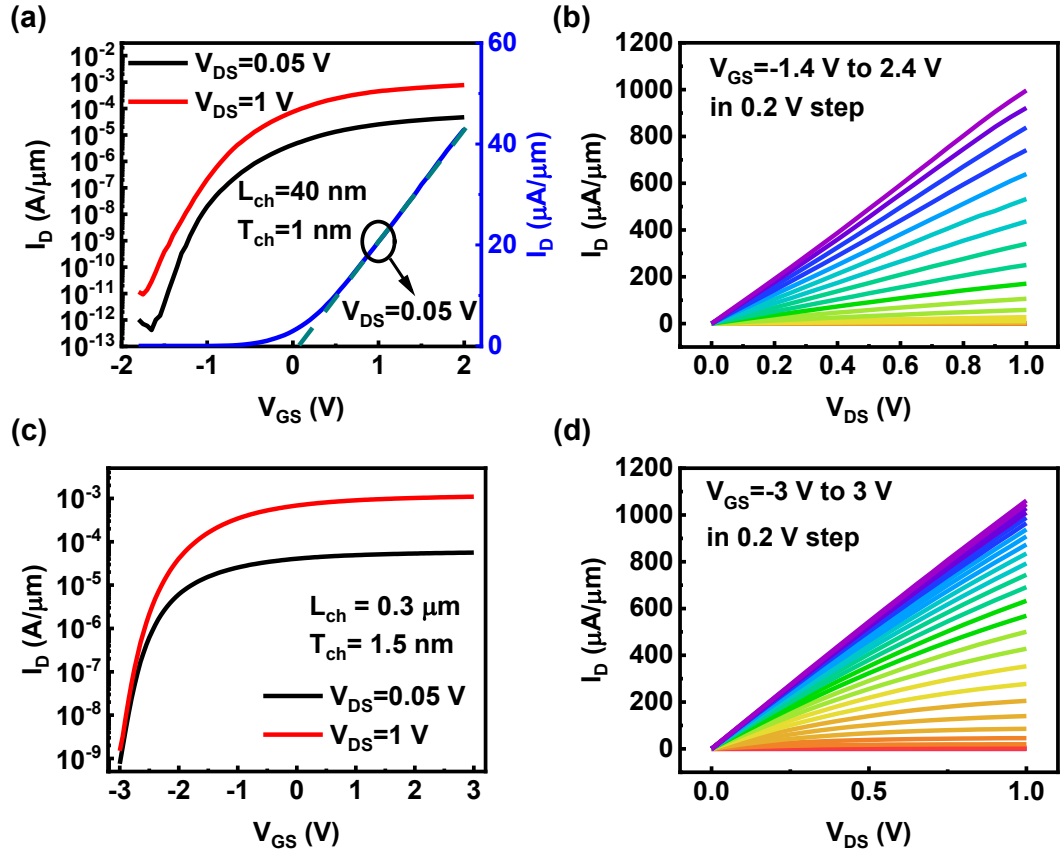

Fig. 4. (a) $I_D$-$V_{GS}$ and (b) $I_D$-$V_{DS}$ characteristics of an $In_2O_3$ transistor with $L_{ch}$ of 40 nm and $T_{ch}$ of 1 nm. (c) $I_D$-$V_{GS}$ and (d) $I_D$-$V_{DS}$ characteristics of an $In_2O_3$ transistor with $L_{ch}$ of 0.3 μm and $T_{ch}$ of 1.5 nm.

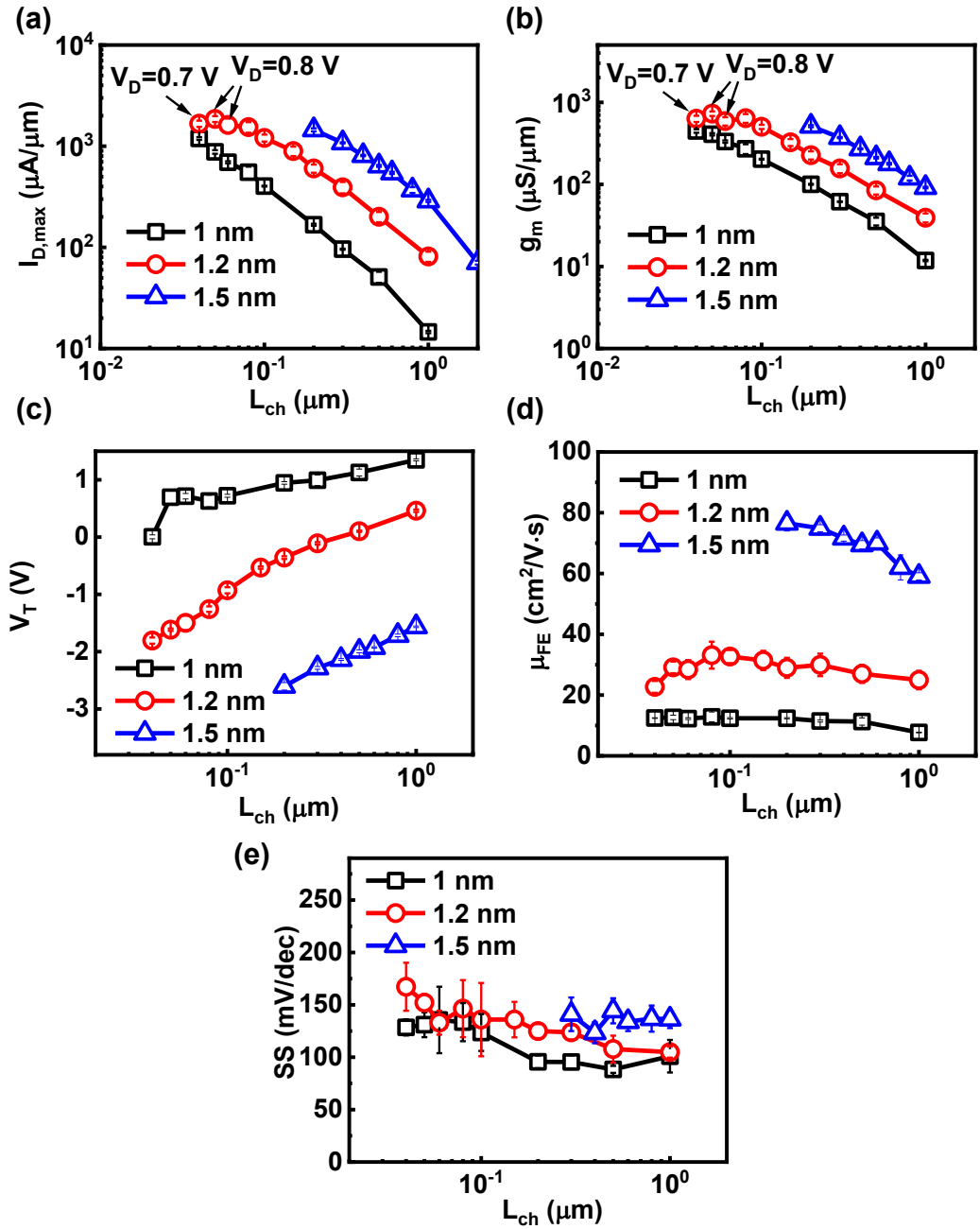

Fig. 5. (a) $I_{D,max}$, (b) $g_m$, (c) $V_T$, (d) $\mu_{FE}$, and (e) SS scaling metrics of $In_2O_3$ transistors with $L_{ch}$ from 1 μm to 40 nm and $T_{ch}$ from 1 nm to 1.5 nm. $I_{D,max}$ and $g_m$ are extracted at $V_{DS}=1$ V unless otherwise specified. Each data point represents the average of at least 5 devices.

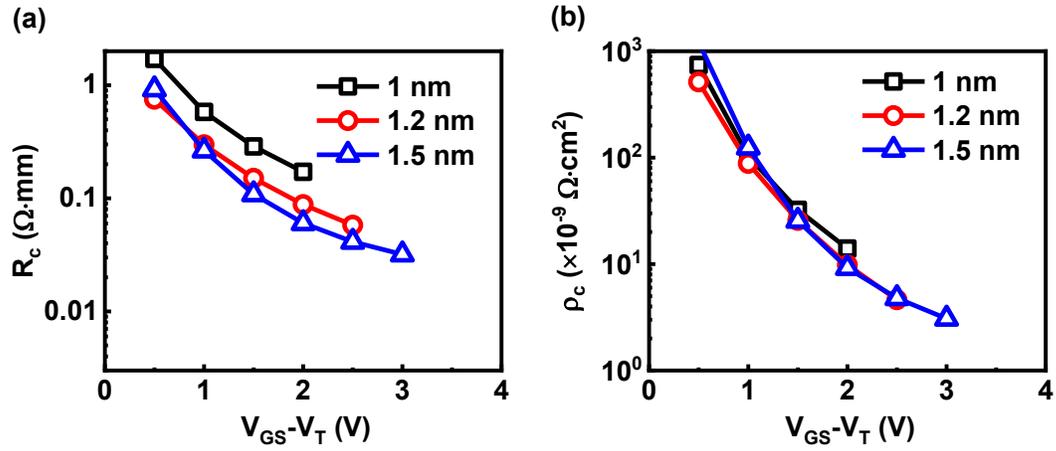

Fig. 6. (a) $R_C$ versus $V_{GS}$-$V_T$ extracted by TLM method for $In_2O_3$ transistors with $T_{ch}$ from 1 nm to 1.5 nm. (b) $\rho_C$ versus $V_{GS}$-$V_T$ extracted by TLM method for $In_2O_3$ transistors with $T_{ch}$ from 1 nm to 1.5 nm.